\documentclass[aps,pra,reprint,superscriptaddress,longbibliography]{revtex4-1}

\usepackage[dvips]{graphicx} 
\usepackage{enumerate}
\usepackage{amsthm,amsfonts,amssymb,amscd,amsmath}
\usepackage{framed}

\newcommand{\ket}[1]{\mbox{$\left| #1 \right\rangle$}}

\begin{document}

\title{Source-independent quantum random number generation}

\author{Zhu Cao} 
\author{Hongyi Zhou}
\author{Xiao Yuan}
\author{Xiongfeng Ma}
\email{xma@tsinghua.edu.cn}
\affiliation{Center for Quantum Information, Institute for Interdisciplinary Information Sciences, Tsinghua University, Beijing 100084 China}

\begin{abstract}
Quantum random number generators can provide genuine randomness by appealing to the fundamental principles of quantum mechanics. In general, a physical generator contains two parts---a randomness source and its readout. The source is essential to the quality of the resulting random numbers; hence, it needs to be carefully calibrated and modeled to achieve information-theoretical provable randomness. However, in practice, the source is a complicated physical system, such as a light source or an atomic ensemble, and any deviations in the real-life implementation from the theoretical model may affect the randomness of the output. To close this gap, we propose a source-independent scheme for quantum random number generation in which output randomness can be certified, even when the source is uncharacterized and untrusted. In our randomness analysis, we make no assumptions about the dimension of the source. For instance, multiphoton emissions are allowed in optical implementations. Our analysis takes into account the finite-key effect with the composable security definition. In the limit of large data size, the length of the input random seed is exponentially small compared to that of the output random bit. In addition, by modifying a quantum key distribution system, we experimentally demonstrate our scheme and achieve a randomness generation rate of over $5\times 10^3$ bit/s.
\end{abstract}

\maketitle

\section{Introduction}
Random numbers play important roles in many fields, such as  scientific simulation \cite{metropolis1949monte}, cryptography \cite{Shannon:1949:OTP},  testing fundamental principles of physics \cite{bell1964einstein}, and lotteries.
Different applications require different levels of randomness.
In  cryptography,  input  randomness  is  one of the  security  foundations in  communication  protocols.  In  fact,  many
commercial products  for generating random  numbers  exist in  the  market;  such  products function under  various
information-theoretical or computational assumptions.

In computer science, random number generators (RNGs) are based on generating pseudorandom numbers \cite{knuth2014art} in which  a random seed is expanded according to some deterministic procedure. By definition, these RNGs produce sequences that are not truly random. Although these sequences usually attain a perfect balance between 0s and 1s, strong long-range correlations exist which undermine cryptographic security and may cause unexpected errors in scientific simulations.

In contrast, hardware RNGs originating from physical processes, such as noise in electric devices, nuclear fission, and circuit and radial decay \cite{kelsey2004entropy,schindler2003evaluation,jun1999intel,holman1997integrated,fort2003very,bucci2003high,tokunaga2008true}, are believed to be able to offer better random numbers. However, it is unclear whether they are truly random because these RNGs normally involve complicated classical physics processes that produce no randomness.

To solve this problem, the new field of quantum random number generators (QRNGs) has emerged. These generators stem from the uncertainty principle in quantum mechanics and are therefore inherently random. Existing QRNG methods include single photon detection \cite{Jennewein:QRNG:2000,Dynes:QRNG:2008,wayne2010low,furst2010high}, vacuum state fluctuations \cite{Gabriel:QRNG:2010}, and quantum phase fluctuations \cite{Xu:QRNG:2012}. These approaches have developed to the point that some commercial QRNG products are available \cite{IDQ,furst2010high,pQRNG150, quRNG, wahl2011ultrafast}.

\begin{figure}[htb]
\centering \resizebox{6cm}{!}{\includegraphics{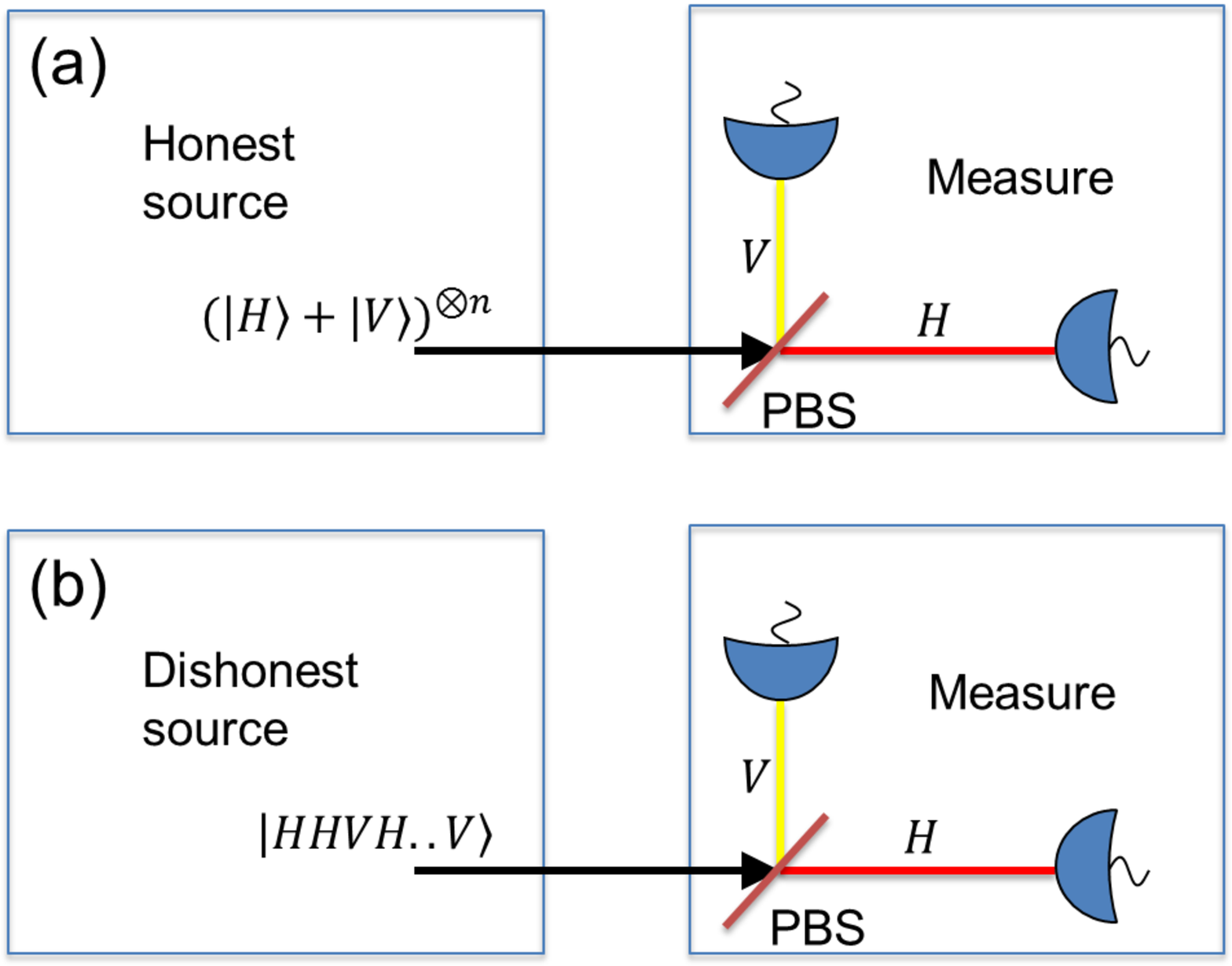}}
\caption{Illustration of a generic QRNG setup in which we take photon polarization as the example. $H$ and $V$ refer to horizontal and vertical polarizations, respectively. PBS refers to a polarizing beam splitter. (a) The source functions normally (or trusted) and sends superpositions of  $H$ and $V$ polarizations, which offers quantum randomness. (b) The source malfunctions (or untrusted) and sends $H$ and $V$ polarizations in a predetermined order, which should output no genuine randomness. From the measurement result viewpoint, one cannot distinguish these two cases.} \label{Fig:SRModel}
\end{figure}

A typical QRNG can be decomposed into two modules: a  randomness source (quantum state preparation) and its readout (measurement), as shown in Fig.~\ref{Fig:SRModel}. In general, the source emits quantum states that are superpositions of the measurement basis. The output (raw) random numbers are the measurement results. In many QRNGs, a short random seed is required to assist state preparation or measurement.

As an example, consider a simple QRNG that projects the quantum state  $\ket{+}=(\ket{H}+\ket{V})/{\sqrt{2}}$ emitted from a single photon source on the horizontal and vertical polarization basis {\ket{H}, \ket{V}}. This QRNG can be divided into two modules, as shown in Fig.~\ref{Fig:SRModel}(a). Randomness is guaranteed by the intrinsically probabilistic nature of quantum physics. Hereafter, we denote \ket{H}, \ket{V} ($\ket{+}$, $\ket{-}$) as the $Z$-basis ($X$-basis) eigenstates.

Existing practical QRNGs suffer from security loopholes if the devices are not perfect.
In the source readout model, the measurement device can normally be trusted due to its simple structure. For instance, in the previous example, the measurement is a simple demolition measurement on the polarization basis.
In contrast, the randomness contained in a source, such as a laser or an atomic ensemble, is normally difficult to characterize completely. If the source malfunctions and emits classical signals instead of quantum ones, the outputs may not be truly random. Consider the worst-case scenario in which the devices are designed or controlled by an adversary Eve. Eve can employ a pseudo-RNG to output a fixed (from Eve's viewpoint) string that still appears random to Alice. More concretely, in the example of the previous paragraph, when a dishonest source emits $Z$-basis instead of $X$-basis eigenstates for the $Z$-basis measurement, the output will just be a fixed string, as shown in Fig.~\ref{Fig:SRModel}(b). From this perspective, with given measurement devices, justifying the randomness in a source is crucial to generating  randomness.

Most existing QRNGs use complicated physical models \cite{PhysRevA.75.032334,Xu:QRNG:2012} to quantify their sources. For example, the dimension of the source is sometimes assumed to be a fixed known number \cite{PhysRevA.90.052327}. The underlying models implicitly assume the existence of randomness in the first place, but this assumption cannot be verified experimentally. Therefore, to achieve truly reliable randomness, there is a strong motivation to avoid the use of such models. Note that removing the dimension assumption is the key challenge to the analysis for device-independent scenarios.

Thus, a QRNG without trusting the source (source-independent) is both theoretically and practically meaningful and greatly  needed. A device-independent QRNG \cite{vazirani2012certifiable} can generate randomness without having to  trust the devices. This type of QRNG requires a short seed for device testing, which is the reason why they are also called randomness expansions \cite{colbeck2011private,miller2014robust,miller2014universal}. By observing the violation of a certain Bell's inequality, such as the Clauser-Horne-Shimony-Holt inequality \cite{CHSH}, one can guarantee the presence of randomness without any assumptions about the source or the measurement device. The main drawback of device-independent QRNGs is that they are not loss tolerant, which typically imposes very severe requirements on experimental devices. Furthermore, this type of QRNG generates random numbers at rates that are very low for practical applications. The highest speed of this type of QRNG has, so far, been reported to be 0.4~bps \cite{PhysRevLett.111.130406}.

Here, we propose a source-independent QRNG (SIQRNG) scheme that is loss tolerant and hence highly practical. In particular, our scheme allows the source to have arbitrary and unknown dimensions. The loss-tolerance property enables potential high-loss implementations of our scheme, such as in integrated optic chips or with inefficient but cheap single photon detectors. We analyze the randomness of the scheme based on complementary uncertainty relations. Our analysis takes into account several practical issues, including finite-key-size effects, multiphoton components in the source, initial seed length, and losses. The analysis combines several ingredients from the security proof of quantum key distribution (QKD), a rich subject that has developed over the last two decades. These ingredients include phase error correction \cite{ShorPreskill_00}, random sampling \cite{Fung:Finite:2010}, and the squashing model \cite{BML_Squash_08}. Since the squashing model shows the equivalence between threshold detectors and qubit detectors \cite{BML_Squash_08}, our scheme allows the source to have an unfixed finite dimension as well as an infinite dimension. For simplicity, in the rest of the paper, we assume a two-level (bit) output system. All our techniques can be directly applied to cases with more outputs.

In many theoretical aspects, there are strong similarities between QKD and QRNG. For example, the security definition in QKD can be applied to the definition of randomness in QRNG, and similar proof techniques can be applied to both, as we do in the later analysis. However, in some practical scenarios, there are subtle differences between the two. For example, local randomness is free in QKD but not in QRNG. A more crucial difference lies in the trustworthy components of QKD and QRNG in practice. In QKD, the sender and receiver are two remotely separated parties, so an adversary could intercept and resend the transmitted signals in the quantum channel and then take advantage of the imperfections of measurement devices to perform attacks. Thus, compared to the source, the measurement device becomes a more vulnerable part of a QKD system.

Different from QKD, source and measurement devices in QRNG are normally local, so attacks aimed at  imperfections in measurement devices seem more artificial than practical. The main purpose in studying the untrusted device scenario in QRNG is to address device imperfections. This subtle difference may lead to deviations between QKD and QRNG. For instance, it is reasonable to assume that Alice can characterize the measurement device for QRNG well and trust it during random number production. Furthermore, compared to QKD, the source in QRNG  involves a complicated design so that the QRNG is fast and convenient. For instance, in a recent experiment \cite{sanguinetti2014quantum}, a QRNG was demonstrated based on measuring  light-emitting  diode  (LED) light with a mobile phone. Such sources are hard to characterize and could possibly be manipulated by Eve, but one can reasonably trust one's own mobile phone. From this viewpoint, the source in QRNG is at least as problematic as the measurement. Thus in our work, we take the reasonable assumption that the measurement device can be characterized well but not the source. Note that the opposite scenario, where the source rather than the measurement device of QRNG is trusted, has also been recently investigated \cite{MDIQRNG15}.

To show the practicality of the proposed scheme, we provide a proof-of-principle experimental demonstration by simply modifying a QKD system. We experimentally examine the effect of different detector efficiencies on the randomness generation rate. Under a practical total transmittance, a high randomness generation rate can be achieved.

The organization of the paper is as follows. In Section~\ref{protocol}, we present our protocol. In Section~\ref{analysis}, we analyze the protocol and calculate the min-entropy of its output after investigating various practical scenarios.
In Section~\ref{sec:exp}, an experimental demonstration of our protocol is performed.
Finally, we conclude in Section~\ref{discussion}.

\section{Protocol}\label{protocol}
\begin{figure}[hbt]
\centering \resizebox{6cm}{!}{\includegraphics{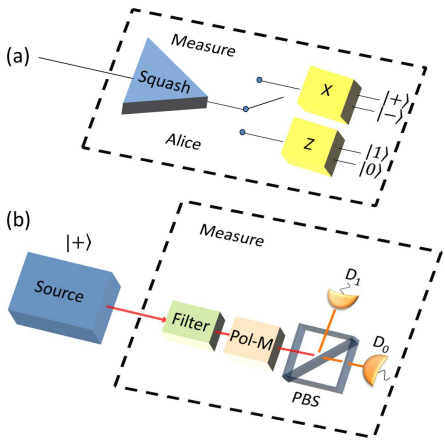}}
\caption{(a) Measurement model for  SIQRNG. The quantum state first passes through a squasher and is projected as either a qubit or a vacuum. Then, the output qubit is measured in the $X$ or $Z$ basis chosen by an active switch. There are two outcomes for each basis measurement, corresponding to the two eigenstates of the basis. (b) An optical implementation of the SIQRNG in (a), as discussed in Section~\ref{Sec:Squash}. Here Pol-M refers to a polarization modulator, PBS refers to a polarizing beam splitter, and $D_0$ and $D_1$ are the threshold detectors.} \label{Fig:EXPSetup}
\end{figure}

A schematic of our SIQRNG protocol is shown in Fig.~\ref{Fig:EXPSetup}(a). Here, we take an optical implementation as the example, as shown in Fig.~\ref{Fig:EXPSetup}(b). All our results apply similarly to other implementation systems. Quantum signals from the source first go through a modulator that actively chooses between the $X$ and $Z$ bases. Then, a polarizing beam splitter and two threshold detectors perform a projective measurement. Since two detectors are used, there are four possible outcomes: no clicks (losses), two single clicks, and double clicks. This implementation is equivalent to the schematic setup of the \emph{squashing model} as discussed in Section \ref{Sec:Squash}. The details of the protocol are presented in Fig.~\ref{Tab:Procedure}.

\begin{table}[htbp]
\begin{framed}
\centering
\begin{enumerate}
\item
\textbf{Source:} An untrusted party, Eve, prepares many quantum states in an arbitrary and unknown dimension and feeds them into the measurement box of Alice.
\item
\textbf{Squashing:} Alice (or Eve) squashes the quantum states into qubits and vacua. Alice postselects  the vacua and obtains $n$ squashed qubits. The vacuum components take into account optical losses and quantum efficiencies.
\item
\textbf{Random sampling:} By consuming a short seed with the length given in Eq.~\eqref{Source:seed}, Alice randomly chooses $n_x$ out of the $n$ squashed qubits and measures them in the $X$ basis, each results in $\ket{+}$ or $\ket{-}$. 
\item
\textbf{Parameter estimation:} When  the  system  operates  properly, the source emits qubits $\ket{+}$ for all runs. Thus, a result of $\ket{-}$ in the $X$-basis measurement is defined as an error. A double click is considered to be  half an error. Alice evaluates the bit error rate $e_{bx}$ in the $X$ basis and its statistical deviation $\theta$ according to Eq.~\eqref{eq:Ptheta}. If $e_{bx}+\theta\ge1/2$, Alice aborts the protocol.
\item
\textbf{Randomness generation:}
For the remaining $n-n_x$ squashed qubits, Alice performs measurement in the $Z$ basis to generate $n_z = n-n_x$ random bits.
\item
\textbf{Randomness extraction:} Alice picks a parameter $t_e$ according to the desired failure probability restriction and extracts $n_z-n_zH(e_{bx}+\theta)-t_e$ bits of final randomness using Toeplitz-matrix hashing \cite{Mansour:Toeplitz:93,Ma2011Finite}\footnote{Other extraction methods, such as Trevisan's extractor \cite{Trevisan:Extractor:1999} can be applied, in which the relation between the failure probability and $t_e$ can differ.}.
\item
\textbf{Security parameter:} With the composable security definition, the security parameter (in trace-distance measure) is given by $\varepsilon=\sqrt{(\varepsilon_\theta+2^{-t_e})(2-\varepsilon_\theta-2^{-t_e})}$.
\end{enumerate}
\end{framed}
\caption{Source-independent QRNG with the finite data size effect. The results are proven in Section~\ref{analysis}.} \label{Tab:Procedure}
\end{table}


\section{Analysis}\label{analysis}
In this section, we  analyze the randomness output of the SIQRNG protocol. Strictly speaking, like device-independent QRNGs, our scheme is  a randomness expansion scheme, in which a random seed is used to generate extra independent randomness. The procedure of parameter estimation is an analog to the phase error rate estimation in QKD postprocessing \cite{Ma2011Finite}. Randomness extraction is mathematically equivalent to privacy amplification in QKD.  The difference between the biased measurement used here and the biased-basis choice QKD protocol \cite{Lo:EffBB84:2005} is that the number of $X$-basis measurements is a constant in our case, whereas in QKD, this number must go to infinity when the data size is infinitely large.

\subsection{Squashing model}  \label{Sec:Squash}
In the SIQRNG scheme, we assume that  measurement devices are trusted and well characterized. The key assumption here is that the \emph{measurement setup is compatible with the squashing model}. In other words, a measurement can be treated in two steps. First, the (unknown arbitrary-dimensional) signal state emitted from the source is projected to a qubit or vacuum. The projection is called  a squasher, as shown in Fig.~\ref{Fig:EXPSetup}(a). Then, the squashed qubits are postselected by discarding the vacua and measuring them in the $X$ or $Z$ basis.  This assumption can be satisfied when threshold detectors are used with random bit assignments for  double clicks \cite{BML_Squash_08}.
For the protocol described in Section \ref{protocol}, the $X$-basis measurement results are used for parameter estimation and  are then discarded in postprocessing. Thus, the random assignment can be replaced by adding  half of the double-click ratio to the $X$-basis error rate.

In practice, it is a challenge to verify whether a measurement setup is compatible with the squashing model. Much effort has been put into this question \cite{PhysRevA.86.042327}. The key point here is to make the two detectors respond equally to (four) different qubits, and hence make the measurement device basis independent \cite{Fung:Mismatch:2009}. This can be done by adding a series of filters (including spectrum and temporal filters) before the threshold detectors, to ensure that the input states stay within a proper set of optical modes \cite{xu2014experimental}, in which the detectors have the same efficiencies \cite{BML_Squash_08, Fung:Mismatch:2009}. One can further assume that Alice uses a trusted source to calibrate the measurement devices beforehand; that is, Alice performs a quantum measurement tomography. A similar measurement calibration procedure should be done in most  current QKD and QRNG realizations. Here, we emphasize that the verification of the squashing model does not affect the source-independent property of our scheme. Thus, we leave detailed investigation on validating the measurement setting for future works.


Similar to the QKD case \cite{BML_Squash_08}, we can assume that the squashing operator is held by Eve in the randomness analysis. By this, we mean that Eve can choose a valid operator, so long as the output is a qubit or a vacuum. In the following discussions, we focus on the squashed qubits. We need to determine the min-entropy associated with these qubits in the $Z$-basis measurement.

\subsection{Complementary uncertainty relation} \label{Sec:Uncertain}

First, we show intuitively why the protocol works. According to quantum mechanics, the outcome of projecting the state $\ket{+}$ on the $Z$ basis is random. Of course, in reality, due to device imperfections, Alice would never obtain a perfect state of $\ket{+}$. Now, the key question for Alice becomes how to verify that the source faithfully emits the state $\ket{+}$. This can be done by borrowing a similar technique from the security analysis of QKD \cite{LoChauQKD_99,ShorPreskill_00,Koashi_Uncer_06} and consider an equivalent virtual protocol depicted in Fig.~\ref{Tab:Procedureeq}, where we replace steps $5$ and $6$ by $5'$ and $6'$. In steps $3$ and $4$ of the protocol, Alice occasionally performs the $X$-basis measurement and defines the \emph{phase error} rate to be the ratio of detecting $\ket{-}$. In the virtual protocol, once Alice knows the phase error rate by random sampling tests, she can perform a phase error correction (step $5'$) before the final $Z$-basis measurement (step $6'$). From the smart design of the phase error correction procedure \cite{ShorPreskill_00}, Alice can make it commute with the $Z$-basis measurement. Thus, she can perform the $Z$-basis measurement (step $5$) first and then apply randomness extraction (step $6$). At this stage, all the states have already collapsed to classical results, and the phase error correction procedure becomes randomness extraction (or privacy amplification in QKD) \cite{LoChauQKD_99,ShorPreskill_00,Koashi_Uncer_06}. Besides QKD, the argument here is similar to the one used in Ref.~\cite{Yuan2015Coherence}, where one can consider the error correction process $5'$ as distilling coherence or randomness extraction.

\begin{table}[htbp]
\begin{framed}
\centering
\begin{enumerate}[1']
\setcounter{enumi}{4}

\item
\textbf{Error correction:} Based on phase error rate $e_{bx}$, Alice performs phase error correction and obtains  $n_z[1-H(e_{bx})]$ copies of perfect $\ket{+}$ with a nearly unit probability.

\item
\textbf{Randomness generation:}
After obtaining all states in $\ket{+}$, Alice performs measurement in the $Z$ basis to get $n_z[1-H(e_{bx})]$ random bits.

\end{enumerate}
\end{framed}
\caption{An equivalent protocol of source-independent QRNG.} \label{Tab:Procedureeq}
\end{table}

It has been proved that the phase error correction (randomness extraction) can be efficiently done with Toeplitz-matrix hashing \cite{Mansour:Toeplitz:93}. Suppose the number of qubits measured in the $Z$ basis is $n_z$ and the phase error rate is $e_{pz}$, the number of bits sacrificed in the phase error correction is given by
\begin{equation} \label{eq:PAcost}
\begin{aligned}
n_zH(e_{pz})+t_e,
\end{aligned}
\end{equation}
and the probability that the phase error correction fails is $2^{-t_e}$ \cite{Ma2011Finite}. Here, $H(e)=-e\log e -(1-e)\log(1-e)$ is the binary Shannon entropy function, all the $\log$ is base 2 throughout the paper, and $t_e$ is the parameter Alice picks up by balancing the failure probability and the final output length. Then, the number of final random bits is given by,
\begin{equation} \label{eq:R}
\begin{aligned}
K\ge n_z-n_zH(e_{pz})-t_e.
\end{aligned}
\end{equation}
In practice, Alice needs to prepare a Toeplitz matrix of size $n_z\times[n_z-n_zH(e_{pz})-t_e]$ for randomness extraction.

We note that the failure probability $2^{-t_e}$ quantifies  fidelity between the  state that results from the phase error correction and the ideal state $\ket{+}^{\otimes n_z}$. In the composable security definition \cite{BenOr:Security:05,Renner:Security:05}, a trace-distance measure security parameter $\varepsilon_t$ should be employed. Its relation to the fidelity measure $\varepsilon_f$ is given by \cite{Fung:Finite:2010}
\begin{equation} \label{eq:traceFidelity}
\begin{aligned}
\varepsilon_t=\sqrt{\varepsilon_f(2-\varepsilon_f)}
\end{aligned}
\end{equation}
In the following, we use the fidelity measure for the failure probability, which, in the end, can be conveniently converted to the trace-distance measure security parameter.

To construct the Toeplitz matrix of size $n_z\times[n_z-n_zH(e_{pz})-t_e]$, Alice needs to use $n_z+n_z-n_zH(e_{pz})-t_e-1$ random bits. Thanks to the Leftover Hash Lemma \cite{Impagliazzo:Leftover:1989}, the Toeplitz hashing extractor can be proven to be a strong extractor. That is, the output random bits are independent of the random bits used in the construction of the Toeplitz matrix \cite{frauchiger2013true}. Thus, the Toeplitz matrix can be reused.

Our result can also be derived via a different but  elegant approach by employing a newly developed seminal uncertainty relation \cite{tomamichel2012tight} and extending the Leftover Hash Lemma \cite{Impagliazzo:Leftover:1989} to the quantum scenario \cite{tomamichel2011leftover}. Interestingly, the result from that approach yields a security parameter (in trace-distance measure) that is of the order of $2^{-t_e/2}$, which is consistent with ours. Such techniques have been successfully applied in some applications, including QRNGs \cite{PhysRevA.90.052327}.

%
%
%
%

\subsection{Finite key analysis} \label{Sec:Finite}
In practice, the QRNG only runs for a finite  time; consequently, the sampling tests for the $X$-basis measurements will suffer from statistical fluctuations. In the parameter estimation step, the key parameter $e_{pz}$ in Eq.~\eqref{eq:R} should be estimated (bounded) from the finite data size effect. 

In the random sampling test, Alice measures the squashed qubits in the $X$ basis and obtains the error rate, $e_{bx}$. Remember that, as required in the squashing model, this error rate includes half of the double-click ratio. Henceforth, we simply call this error rate the $X$-basis error rate. Recall that the phase error rate $e_{pz}$ is defined as the error rate if the quantum signals measured in the $Z$ basis are measured in the $X$ basis. When the sampling size is large enough, $e_{pz}$ can be well approximated by $e_{bx}$. 

Before presenting the details of the random sampling analysis, we establish a notation. Suppose Alice receives $n$ squashed qubits and randomly chooses $n_x$ of them to be measured in the $X$ basis, leaving the remaining $n_z=n-n_x$ qubits in the $Z$ basis. Let the ratio of $X$-basis measurements be $q_x=n_x/n$, the number of errors Alice finds in the $X$ basis be $k$, and the total number of errors be $m$ if Alice measures all qubits in the $X$ basis. Then, the number of errors in the qubits measured in the $Z$ basis is $m-k$, which is the key parameter we need to determine through random sampling. The quantity $m-k=n_ze_{pz}$ determines the randomness extraction rate. Define the lower bound of $e_{pz}$ by,
\begin{align} \label{eq:eptheta}
e_{pz}\le e_{bx}+\theta,
\end{align}
where $\theta$ is the deviation due to statistical fluctuations.
Following the random sampling results of Fung et al.~\cite{Fung:Finite:2010}, we can bound the probability when Eq.~\eqref{eq:eptheta} fails,
\begin{equation} \label{eq:Ptheta}
\begin{aligned}
\varepsilon_\theta &= \text{Prob}(e_{pz}>e_{bx}+\theta)  \\
&\le \frac{1}{\sqrt{q_x(1-q_x)e_{bx}(1-e_{bx})n}}2^{-n\xi(\theta)},
\end{aligned}
\end{equation}
where $\xi(\theta)= H(e_{bx}+\theta-q_x\theta)-q_x H(e_{bx})-(1-q_x)H(e_{bx}+\theta)$. Note that in the unlikely event that $e_{bx}=0$, the failure probability is unbounded, and one should rederive the failure probability or simply replace $e_{bx}$ with a small value, say, $1/n_x$.

In practice, the failure probability $\varepsilon_\theta$ is normally picked to be a small number depending on applications. In later data postprocessing, we pick up $\varepsilon_\theta=2^{-100}$. Once $\varepsilon_\theta$ is fixed, there is a trade-off between $q_x$ and $\theta$ for the ratio of the final random bit length over the raw data size. Thus, the number of samples for the $X$-basis measurement should be optimized for the randomness extraction rate.

One key property for the random sampling is that the $n_x$ locations of the $X$-basis measurements are randomly chosen from the total $n$ locations, i.e., the $\binom{n}{n_x}$ cases are equally likely to occur. Then, Alice needs a random seed with a length of
\begin{equation} \label{Source:seed}
n_{seed} = \log\binom{n}{n_x} \le n_x\log n.
\end{equation}
The effect of loss on the seed length will be discussed in Section \ref{Sec:Issues}.
In Appendix \ref{App:numXmeas}, we show that $n_x$ can remain a constant, given the failure probability, when $n$ is large. Then, in the large data size limit, the seed length is exponentially small compared to the length of the output random bit. Therefore, we reach an exponential randomness expansion.

\subsection{Practical issues} \label{Sec:Issues}
{\bf Multiphotons:} In our protocol, the source is allowed to emit multiphotons, since its dimension is assumed to be uncharacterized. In other words, these components do not affect the randomness of the final output. In practice,  multiphotons may introduce double clicks when threshold detectors are used \cite{BML_Squash_08}; these  double  clicks will directly contribute to the error rate term $e_{bx}$. Thus, when the multi-photon ratio is very high, the double-click ratio will increase to a point at which the upper bound on information leakage $e_{pz}$ increases to one-half; at that point, no random bits can be extracted according to Eq.~\eqref{eq:R} and Alice simply aborts the protocol.


{\bf Loss:} The loss tolerance of our protocol is guaranteed by the squashing model in which the measurement is assumed to be basis independent \cite{BML_Squash_08}. This assumption can be guaranteed by the fact that the basis is chosen after losses. Alice does not anticipate the positions of losses, so she effectively decides the (random) positions for $X$-basis measurements before losses. The effect of loss only decreases the number of effective $X$ measurements, but the positions of effective $X$ measurements are still uniformly random in squashed qubits; this fulfills the requirement of  random sampling. The detailed proof is shown in Appendix \ref{app:sampling}.

{\bf Basis-dependent detector efficiency:} Our protocol  assumes that the efficiencies of the detectors are the same. In practice, efficiency mismatches would cause the measurement to be different for the two bases (basis dependent). A viable way to deal with this imperfection is to recalculate the rate as a function of the ratio between the efficiencies of the two bases, employing the technique used in QKD \cite{Fung:Mismatch:2009}. As indicated by the result in QKD \cite{Fung:Mismatch:2009}, the random number generation rate will slightly decrease when there is a small mismatch in detector efficiencies. More precisely, we denote the ratio between the minimum and maximum efficiencies of the two detectors as $r\le1$, then the key size becomes $r n_z (1-H[(e_{bx}+\theta)/r])-t_e$ bits. We leave detailed analysis of this imperfection for future work.


{\bf Double clicks:} Our analysis takes account of the effect of double clicks by adding half of the double-click ratio to the $X$-basis error rate, as required in the squashing model. This is also essentially why multiphoton states can be used on the source side without affecting final randomness. Note that double clicks should not be discarded freely in the measurement. Otherwise, a security loophole will appear, namely, a strong pulse attack \cite{Lutkenhaus_99DoubleClick}. In a strong pulse attack, Eve always sends strong signals (with many photons) in the $Z$ basis. Suppose she sends a strong state in $\ket{H}$; if Alice chooses the $Z$-basis measurement, a valid raw random bit will be obtained, but if she chooses the $X$ basis, a double click is likely to happen. In our protocol, when Alice chooses the $X$-basis measurement, she should get an error (resulting in $\ket{-}$) with a probability of one-half. If Alice simply discards all double clicks, Eve's attack will not be noticed. This attack cannot be explained by a qubit measurement. This is intuitively why the squashing model requires random assignments for double clicks.


{\bf Basis choice:} 
When choosing $X$- or $Z$-basis measurements, an input random string of length $N$ (as a seed) is needed. Suppose the number of $X$-basis measurements to be performed is $N_x$, then Alice chooses $N_x$ positions out of $N$ with equal probability, i.e., with probability $\binom{N}{N_x}^{-1}$.  Then, she needs a seed length of $\log\binom{N}{N_x}$. This is similar to Eq.~\eqref{Source:seed} with the difference  that before the measurement, Alice does not know the positions of losses. More details on how to dilute a short random seed to a longer (partially random) one are provided in Appendix \ref{app:input}.

{\bf Intensity optimization:} The intensity of the source should be optimized to maximize the randomness generation rate. With increasing intensity, the detection rate will increase along with an increases in the double-click rate (and hence $e_{pz}$ increases). There exists a trade-off between $n_z$ and $e_{pz}$, as shown in Eq.~\eqref{eq:R}.

\section{Experiment demonstration} \label{sec:exp}
\begin{figure}[htb]
\centering \resizebox{9cm}{!}{\includegraphics{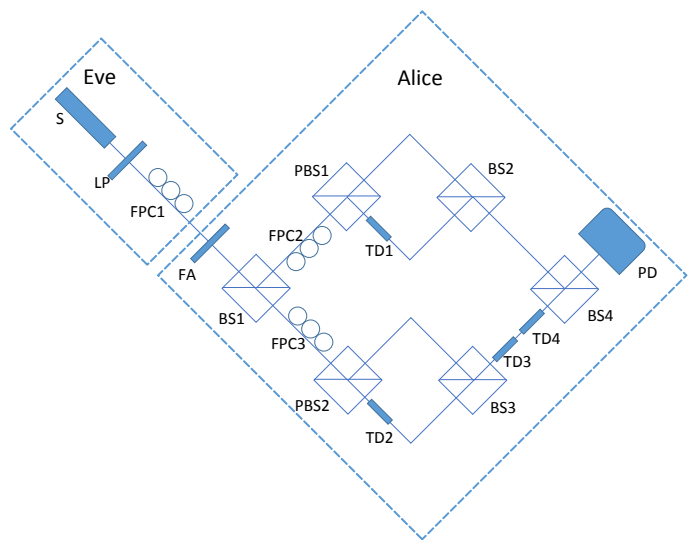}}
\caption{Experiment setup of SIQRNG. S: laser source; LP: linear polarizer; FPC: fiber polarization controller; FA: fiber attenuator; BS: beam splitter; PBS: polarizing beam splitter; TD: time delay implemented with a 12 m fiber; PD: photon detector.}
\label{fig:expsetup}
\end{figure}

In this section, we perform a proof-of-principle experimental demonstration to show the practicality of the SIQRNG scheme. Our experiment setup consists of two parts, the source, owned by an untrusted party Eve, and the measurement device, owned by the user Alice. The schematic diagram is shown in Fig.~\ref{fig:expsetup}.

On Eve's side, a laser, labeled as $S$, with a wavelength of 850 nm and a repetition rate of 1 MHz is used as a photon source. The power of the laser is adjusted to be one photon per pulse. Instead of assuming each state is a qubit system, each pulse that the laser sends is a coherent state of infinite dimensions.  The pulse of the laser is then modulated to $\ket{+}$ polarization by a linear polarizer (LP) and a fiber polarization controller (FPC1). Between the source and the measurement device, we put a fiber attenuator (FA) to simulate different losses in the system.

On Alice's side, first a series of filters needs to be applied to ensure the measured optical mode is pure before entering the threshold detectors, as required by the squashing model. For demonstration purposes, we use a single-mode fiber to play the role of a filter. Ideally, frequency and temporal filters should also be added to further purify the optical mode in order to make the photons indistinguishable. For demonstration purposes, a biased beam splitter (BS1) with a ratio of $1:49$ is used to passively choose the $X$ or $Z$ basis. 
Finally, Alice records when the photon detector (PD) clicks. The detector is time-division-multiplexed by adding four time delays TD1--TD4 (60~ns each) in the optical paths, so that it can simulate four detectors that detect the outcomes of both bases and each bit value. The gate width and the dead time of the detector are 10 ns and 50 ns, respectively.

\begin{figure}[htb]
\centering \resizebox{8cm}{!}{\includegraphics{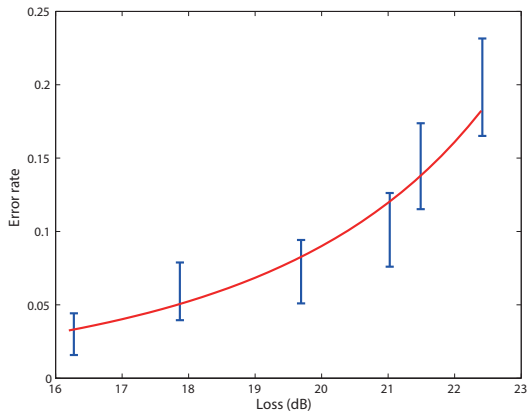}}
\caption{Relation between the phase error rate and the loss. The big error bars are caused by a very conservative estimation of statistical fluctuations and also partially by the fluctuation of experimental parameters for different losses.}
\label{fig:errorrate}
\end{figure}

The phase error rate, as calculated in Eq.~\eqref{eq:eptheta}, is plotted in
Fig.~\ref{fig:errorrate}. The typical values of the related experimental parameters are listed as follows. The raw key size is $N=10^6$; the dark count is $0.002$; the detector efficiency (without a FC adaptor) is $45\%$; the misalignment error of the source is 2\%; and the failure probability is $\varepsilon_\theta=2^{-100}$.  The figure shows that the error rate increases as the loss becomes large. This is because the effect of dark counts becomes dominant when the loss is high. Because of statistical fluctuations, the phase error rate increases when the data size shrinks. Note, in particular, that the phase error rate can go beyond $20\%$ under high losses, which does not yield any key rates in most QKD protocols. Nevertheless, random numbers can still be generated in our SIQRNG scheme.


\begin{figure}[htb]
\centering \resizebox{8cm}{!}{\includegraphics{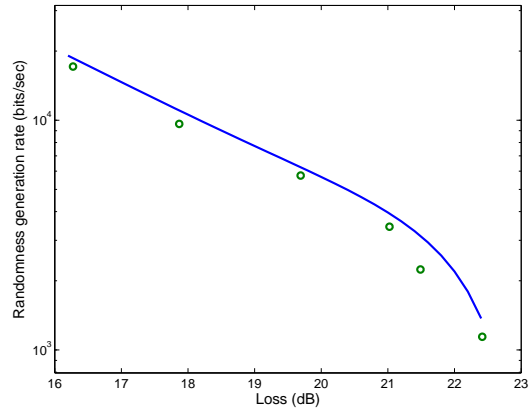}}
\caption{Dependency of randomness generation rate on the loss. The data points on the figure are taken to be the lower bound of the rate, evaluated by random sampling. The security parameter is $\varepsilon_t=2\times 2^{-50}$}
\label{fig:rngrate}
\end{figure}

The relation between the randomness generation rate and the loss is plotted in Fig.~\ref{fig:rngrate}.  It can be seen that the randomness generation rate becomes lower with a larger loss, which is consistent with Fig.~\ref{fig:errorrate}. Under practical detector efficiency, the randomness generation rate still achieves a relatively high rate of $5\times 10^{3}$~bit/s. Note that, the intensity of the source is fixed in our experimental demonstration. In practice, the intensity of the source can be increased to compensate the loss, and actually the maximum randomness generation rate in our scheme is mainly limited by the dead time of the detector. For our detector with a dead time of 50 ns, the maximum randomness generation rate is $1$ bit$/50$ ns=20 Mbps, which requires the source to be a single photon source with a repetition rate of 20 Mbps. For practical implementations with coherent-state sources, the randomness generation rate can reach the order of 2 Mbps after taking into account various errors and finite data size effects.


\begin{figure}[htb]
\centering \resizebox{9cm}{!}{\includegraphics{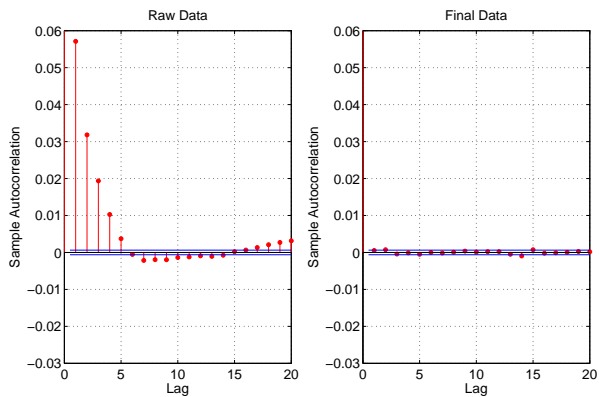}}
\caption{The autocorrelation function of the raw data and the final data. The $x$ axis is the lag $j$ between the sampled data $X_i$ and $X_{i+j}$, while the $y$ axis is the autocorrelation $R(j)$ defined in Eq.~\eqref{eq:auto}. Data sizes of both the raw data and the final data are on the order of $10^7$. The autocorrelation of the final data is significantly smaller than the raw data in absolute value. Because of finite-key-size effects, the autocorrelation cannot be zero even for perfectly random strings.}
\label{fig:AR}
\end{figure}

\begin{figure}[htb]
\centering
\includegraphics[width=8cm]{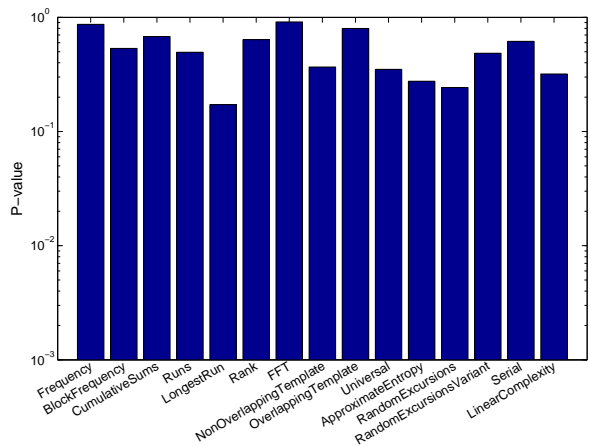}
\caption{The $P$ value of the statistical tests. The $x$ axis lists the names of statistical tests in the NIST test suite. The final data size is 91 Mbit, which is extracted from 115-Mbit raw data. To pass each test, the $P$ value should be at least 0.01 and the proportion of sequences that satisfy $P>0.01$ should be at least 96\%. It can be seen  in the figure that the $P$ values of all tests are greater than $0.01$.}
\label{fig:p}
\end{figure}

After obtaining the random bits, we apply the Toeplitz-matrix hashing \cite{Mansour:Toeplitz:93} on the raw data to obtain final random numbers. To test the randomness, we further perform two statistical tests on the output of our SIQRNG, the autocorrelation test and the NIST test suite \cite{NIST}.  The autocorrelation is defined as
\begin{equation}\label{eq:auto}
 R(j) = \frac{ \mathbb{E}[ (X_i-\mu) ( X_{i+j}- \mu )]}{\sigma^2},
\end{equation}
where $j$ is the lag between the samples, $X_i$ is the $i$-th sample bit, $\mu$ and $\sigma$ are the average and the variance of the sample, and $\mathbb{E}$ stands for expectation. The result of the autocorrelation test of raw data and final data is shown in Fig.~\ref{fig:AR}. It can be seen that the autocorrelation is substantially reduced in the final data.
The result of NIST tests on the final data is shown in Fig.~\ref{fig:p}. We can see that all tests are passed.

\section{Conclusion} \label{discussion}
We have proposed a source-independent and loss-tolerant QRNG scheme and its experimental demonstration in a passive basis choice realization. From an experimental point of view, the beam splitter itself, as part of the measurement device, may also be uncharacterized. Thus, it would also be interesting to demonstrate our scheme with an active basis choice in the future. In fact, when the source operates properly, the speed of our protocol is comparable to that of a trusted polarization-based QRNG whose frequency is limited only by single photon detectors---approximately 100~Mbps \cite{comandar2014ghz}.

Some current realizations of QRNG experiments could be converted to our SIQRNG protocol. For example, a LED could be used as the source, as  regular QRNG \cite{sanguinetti2014quantum}. Since the polarizations of a LED are random, it would be convenient to add a polarizer for the $\ket{+}$ direction to make the source-polarized light. Since the detector can work in a gated mode, it does not matter whether the light source is  continuous or pulsed. This shows why the repetition rate is limited only by single-photon detectors. Viewed from another angle, such a setup could also be used to test  quantum features of macroscopic sources.


For future projects, it would be interesting to investigate other loss-tolerant self-testing QRNG schemes. Essentially, we are aiming to design a QRNG to tolerate large losses and generate fast random numbers simultaneously, given the minimum assumptions of a practical setup.

{\it Added note:} Upon completion of this work, we noticed a related work \cite{PhysRevA.90.052327}, the uncertainty relation is employed to quantify entropy in QRNG and finite-key effects are taken into consideration with smooth min entropies. The work also aimed at provable randomness with untrusted sources. However, it makes a strong assumption on the dimension of the source, which turns out to be the key barrier for source-independent QRNG. Moreover, the practical imperfections, such as multi-photons, device imperfections and losses, are not considered. Our work, on the other hand, use the squashing model for arbitrary dimension system and take account of imperfections in practical scenarios.




\section*{Acknowledgements}
The authors acknowledge insightful discussions with T.-Y.~Chen, H.-K.~Lo, N.~L\"utkenhaus, B.~Qi, R.~Renner, Y.~Shi, F.~Xu, and Z.~Yuan, and especially with Y.-Q.~Nie on the experiment and randomness tests. The authors also acknowledge experiment devices supported by QuantumCTek Co., Ltd. This work was supported by the National Basic Research Program of China Grants No.~2011CBA00300 and No.~2011CBA00301 and the 1000 Youth Fellowship program in China.

Z.~C.~and H.~Z.~contributed equally to this work.

\appendix
\section{Calculation of the number of effective $X$-basis measurements}
\label{App:numXmeas}
In this appendix, we show that in the asymptotic limit, the number of effective $X$-basis measurements is independent of $n$. Our starting point is Eq.~\eqref{eq:Ptheta} and $\varepsilon_\theta < 2^{-100}$. Notice that normally $n$ is smaller than $10^{12} < 2^{40}$ to ease fast postprocessing; thus, the term $1/\sqrt{n}$ and the other polynomial terms in Eq.~\eqref{eq:Ptheta} play a relatively small role in making $\varepsilon_\theta < 2^{-100}$. In the following, we consider only the exponent in Eq.~\eqref{eq:Ptheta}.

For ease of notation, let $x=e_{bx}$, $y=e_{bx}+\theta$ and $q=q_x$. Then the exponent of Eq.~\eqref{eq:Ptheta} becomes
\begin{equation*}
n[H((1-q)y+q x)-q H(x)-(1-q) H(y)]
\end{equation*}
and the inequality $\varepsilon_\theta < 2^{-100}$ is approximately equivalent to
 \begin{equation}
 \begin{aligned}
& n [q(H((1-q)y+q x)-H(x))+ \\
 & (1-q) (H((1-q)y+q x )- H(y))] \ge 100.
  \end{aligned}
\label{eq:A1}
 \end{equation}
Since $q$ is very small, one can make three approximations:
\begin{equation} \label{firstapprox}
H ((1-q)y+qx )-H(y)\approx -H'(y)q(y-x),
 \end{equation}
 \begin{equation} \label{secondapprox}
q[H((1-q)y+qx)-H(x)]\approx q(H(y)-H(x))
 \end{equation}
and
\begin{equation} \label{thirdapprox}
q^2 \approx 0.
 \end{equation}
Then, by applying Eqs.~\eqref{firstapprox} and \eqref{secondapprox}, the inequality \eqref{eq:A1} becomes
 \begin{equation}
n[q(H(y)-H(x))-(1-q) (H'(y)q(y-x))] \gtrsim 100.
 \end{equation}
 Applying Eq.~\eqref{thirdapprox} yields
 \begin{equation}
n[q(H(y)-H(x))- H'(y)q(y-x)] \gtrsim 100,
 \end{equation}
 and rearranging terms, we have
 \begin{equation}
q \gtrsim \frac{100}{n[H(y)-H(x)-H'(y)(y-x)]}.\\
 \end{equation}
 Substituting the definitions of $x$ and $y$, we obtain
 \begin{equation}
 q \gtrsim  \frac{100}{n[H(e_{bx}+\theta)-H(e_{bx})-H'(e_{bx}+\theta)\theta]}.
 \end{equation}
Finally, we substitute $q=n_x/n$ and get
\begin{equation} \label{eq:c}
n_x \approx \frac{100}{H(e_{bx}+\theta)-H(e_{bx})-H'(e_{bx}+\theta)\theta},
\end{equation}
which is independent of $n$.

\section{Proof of the random sampling property for a type of QRNG input after loss}
\label{app:sampling}
In this appendix, we first restate the setting. In the idealistic protocol, the measurement device chooses its measurement basis after confirming that the state received from the source is not a vacuum (or equivalently, not lost). In practice, confirming whether a state is
a vacuum is usually done by observing whether detectors in the measurement device click or not. Thus, it is desirable for the measurement device to choose its basis before confirming whether loss happens.

We prove that for a specific input that defines the measurement basis choices before the potential loss, the positions of $n_x$ valid $X$-basis measurements (after excluding loss events) are randomly drawn from the positions of the total of $n$ valid measurements. This proves that the random sampling technique from  Fung {\it et al.} can still be applied  when the measurement basis is chosen before the loss.

For ease of presentation, we state the input that specifies the measurement choices before the loss as follows. The input is a string of length $N=N_x+N_z$ that contains $N_x$ 0s and $N_z$ 1s. The $\binom{N}{N_z}$ possibilities for choosing the positions of $N_z$ 1s from the total $N_x+N_z$ positions are equally likely. Here, 0 stands for an $X$-basis measurement and 1 stands for a $Z$-basis measurement.  After loss, the numbers of valid $X$-basis measurements and $Z$-basis measurements are denoted by $n_x$ and $n_z$, respectively, with a total string length of
\begin{equation}
n=n_x+n_z.
\end{equation}
We need to show that the output is uniform for the $\binom{n_x+n_z}{n_z}$ possibilities of choosing the positions of $n_z$ 1s from the total $n$ positions.

The proof proceeds through a symmetry argument. The input is symmetric, i.e., if we exchange the indices of two positions, the distribution will not change. Suppose that the initial positions are $1,2,\dots, n$ and the probability of choosing specific positions for $N_z$ 1s from the total $N$ positions is
\begin{equation}
p=\frac{1}{\binom{N_x+N_z}{N_z}}.
\end{equation}
 For ease of presentation, denote the left positions after loss as $i_1<i_2< \dots<i_n$. Then each possibility with $n_x$ 0s in the left $n$ positions has the same probability
 \begin{equation}
 p_1=p \times \binom{N-n}{N_x-n_x},
 \end{equation}
which proves our claim.

As a side remark, we see that the proof does not depend on whether the loss is basis dependent or independent. Thus, the same property also holds for a more general class of losses that could be useful in other settings. Another remark is that independent and identically distributed input also satisfies the property, as in the work of Fung {\it et al}.

\section{Random seed dilution}
\label{app:input}
The input is either given directly or expanded from a uniformly random seed. Here, we provide a method for performing the expansion. The expansion is straightforward since the input is also uniformly random within its support. We can simply map a uniform seed of length $\log \binom{N}{c_1}$ bijectively to the input support, which is the $\binom{N}{c_1}$ possibility of choosing the positions of $c_1$ 0s from the string of length $N$. Then, we obtain the desired input. Furthermore, note that this construction is deterministic; thus, input randomness is only needed for the uniformly random seed of length $n$.

For the input of our protocol, the ratio of the initial random seed length to the number of runs $N$ becomes negligible as $N$ goes to infinity because the number of $X$-basis measurements $c_1$ is a constant, as derived in Appendix \ref{App:numXmeas}. More precisely, the min entropy of the input as well as the length of the uniformly random seed has an upper bound given by
\begin{equation}
\log \binom{N}{c_1}\le c_1 \log N.
\end{equation}
Note that since the detector completely controls this random seed length, calculating the exact input min entropy is possible. This is very different from estimating the error rate in the finite-key analysis section, in which we can only estimate the range of the error rate with a high  probability of success. Apart from the input specified in the main text, independent and identically distributed bit strings are also a possible choice  for the input. Finally, we remark that the reason to include this input seed length analysis is to make our QRNG composable.

\bibliography{BibliSemi}

\end{document}